\pdfoutput=1

\documentclass[11pt]{article}

\usepackage[final]{acl}

\usepackage{times}
\usepackage{latexsym}

\usepackage[T1]{fontenc}

\usepackage[utf8]{inputenc}

\usepackage{microtype}

\usepackage{inconsolata}

\usepackage{graphicx}

\usepackage{amsmath, amssymb}
\usepackage{algorithm}
\usepackage{algorithmic}
\usepackage{amsthm}
\usepackage{bm}
\newtheorem{theorem}{Theorem}

\usepackage{booktabs}
\usepackage{pifont}
\usepackage{multirow}
\usepackage{makecell}
\usepackage{subfigure}
\usepackage{array}
\usepackage{tikz}

\title{MorphMark: Flexible Adaptive Watermarking for Large Language Models}

\author{
    Zongqi Wang,
    Tianle Gu,
    Baoyuan Wu\thanks{Yujiu Yang and Baoyuan Wu are co-corresponding authors.},
    Yujiu Yang\footnotemark[1]
    \\
    $^{1}$Tsinghua University \quad $^{2}$The Chinese University of Hong Kong, Shenzhen
    \\
    $^{1}$zq-wang24@mails.tsinghua.edu.cn, gtl23@mails.tsinghua.edu.cn
    \\
    $^{2}$wubaoyuan@cuhk.edu.cn, $^{1}$yang.yujiu@sz.tsinghua.edu.cn
}

\begin{document}
\maketitle


\begin{abstract}
Watermarking by altering token sampling probabilities based on \emph{red-green} list is a promising method for tracing the origin of text generated by large language models (LLMs). However, existing watermark methods often struggle with a fundamental dilemma: improving watermark effectiveness (the detectability of the watermark) often comes at the cost of reduced text quality. This trade-off limits their practical application. 
To address this challenge, we first formalize the problem within a multi-objective trade-off analysis framework. Within this framework, we identify a key factor that influences the dilemma. Unlike existing methods, where watermark strength is typically treated as a fixed hyperparameter, our theoretical insights lead to the development of MorphMark—a method that adaptively adjusts the watermark strength in response to changes in the identified factor, thereby achieving an effective resolution of the dilemma. 
In addition, MorphMark also prioritizes flexibility since it is an model-agnostic and model-free watermark method, thereby offering a practical solution for real-world deployment, particularly in light of the rapid evolution of AI models. 
Extensive experiments demonstrate that MorphMark achieves a superior resolution of the effectiveness-quality dilemma, while also offering greater flexibility and time and space efficiency. 
\end{abstract}

\section{Introduction}

The rapid development and widespread adoption of Large Language Models (LLMs) have raised concerns about the traceability of AI-generated text and copyright protection. Watermarking~\cite{kirchenbauer2023watermark, liu2024survey, dathathri2024scalable}, which embeds distinctive patterns into generated content, has emerged as a critical solution to these challenges. 
However, the trade-off between watermark effectiveness (i.e., detectability and robustness in this paper) and text quality remains a major barrier to practical adoption. A stronger watermark enhances effectiveness but degrades text quality~\cite{kirchenbauer2023watermark, liu2024survey, dathathri2024scalable}, while a weaker watermark preserves text quality but becomes harder to detect and more vulnerable to attacks, even simple paraphrasing~\cite{liu2024survey, dathathri2024scalable, giboulot2024watermax, wu2024resilient}. Therefore, developing a watermarking mechanism that can effectively reconcile watermark effectiveness and text quality is crucial. 

KGW~\cite{kirchenbauer2023watermark} is the first watermarking method based on red-green lists. Specifically, during token generation, it partitions the vocabulary into green and red lists and then increases green tokens' probabilities. As a result, the generated sequence contains more green tokens, allowing it to be identified as watermarked. However, KGW struggles to balance watermark effectiveness and text quality. 
Unbiased watermarking~\cite{kuditipudi2024robust, hu2024unbiased, wu2024resilient, mao2024watermark} ensures that the expected sampling distribution remains unchanged, preserving text quality. However, current implementations often lack robustness. 
Low-entropy watermarking~\cite{lu2024entropy, lee2024wrote, liu2024adaptive} targets low-entropy text generation. While not explicitly designed for quality preservation, it achieves this by avoiding watermarking low-entropy tokens. However, it requires access to the original model for detection, increasing computational cost. 
Besides, some methods~\cite{liu2024sir, he2024xsir, huo2024token} attempt to balance watermark effectiveness and text quality by training auxiliary models. However, these approaches lack flexibility (model-agnostic and model-free). First, they require training model-specific auxiliary models for different LLMs. Second, they disrupt end-to-end inference, increasing the complexity of LLM deployment and increasing inference latency since they adopt extra models. Therefore, in our paper, we argue that the watermark methods should prioritize flexibility. 

In this paper, we first formulate the watermark effectiveness and text quality as a multi-objective trade-off analysis function to analyze the factors influencing this function. The watermark studied here is also based on the green-red list approach. Through this theoretical framework, we reveal that the cumulative probability of green-list tokens plays a key role in determining the overall multi-objective benefits of increasing watermark strength. Note that watermark strength refers to the parameter that indicates the intensity of the watermark, while watermark effectiveness reflects its practical detectability performance. Specifically, as the cumulative probability of the green list decreases, the benefits of increasing watermarking strength diminish progressively and can even turn negative. 
Based on this theoretical insight, we propose MorphMark, which can effectively address the dilemma between watermark effectiveness and text quality. The core idea of MorphMark is to dynamically adjust the watermarking strength in response to changes in the cumulative probability of the green list, aiming to increase the overall multi-objective benefits. 

We summarize our contributions as follows: 
\textbf{1)} We present a theoretical framework that captures both watermark effectiveness and text quality. Based on this framework, we derive and reveal the critical role of the cumulative probability of green-list tokens in balancing watermark effectiveness and text quality. To the best of our knowledge, this is the first time this role has been revealed. 
\textbf{2)} We introduce MorphMark, a novel watermarking framework that dynamically adjusts watermarking strength based on the cumulative probability of green-list tokens. MorphMark is theoretically sound, effectively addressing the dilemma between text quality and watermark effectiveness. It also demonstrates excellent time and space efficiency. Moreover, it is highly flexible, supporting training-free and end-to-end operation. 
\textbf{3)} Through comprehensive empirical evaluation, we demonstrate the effectiveness and flexibility of MorphMark. 

\section{Preliminaries}
\label{sec:preliminary}

Watermark injection aims to embed a detectable pattern into generated text by modifying the probability distribution output by LLMs. We formalize watermarking in LLMs using KGW~\cite{kirchenbauer2023watermark} as an example below. 

Let the vocabulary be denoted as $\mathcal{V}$, and the input token sequence as $\left( x_1, x_2, \dots, x_{t-1} \right) \in \mathcal{V}^*$. The probability distribution for generating the next token $x_t$ without a watermark is given by:

\begin{equation}
    P\left( x_t \mid x_1, x_2, \dots, x_{t-1} \right),
\end{equation}

\noindent which can be simplified as:

\begin{equation}
    P\left( x_t \mid \boldsymbol{x}_{1:t-1} \right),
\end{equation}

\noindent where $\boldsymbol{x}_{1:t-1} = x_1, x_2, \dots, x_{t-1}$ represents the input sequence.

KGW watermark injection operates as follows: A hash value $ h $ is generated using a user-defined private key $k$ and a preceding token ${x}_{t-1}$. This hash value $ h $ serves as a random seed to partition the vocabulary $ \mathcal{V} $ into two subsets: the green list $ G $ and the red list $ \mathcal{V}_R $, where the green list $ \mathcal{V}_G $ contains a fraction $ \gamma $ of the total vocabulary $ \mathcal{V} $, i.e., $ |\mathcal{V}_G| = \gamma |\mathcal{V}| $. $\gamma$ is set to 0.5 below by default. 

Next, KGW increase the probability of tokens in green list. For simplicity, we will only describe the increase in the probability of green-list tokens, while the probability of red-list tokens will naturally decrease accordingly. Specifically, for a token $ i $, the probability $ p_i $ is modified as follows: 

\begin{equation}
\hat{p}_i =
\begin{cases}
\frac{p_i e^{\delta}}{\sum_{j \in G} p_j e^{\delta} + \sum_{j \in R} p_j}, & \mathcal{V}_i \in \mathcal{V}_G, \\
\frac{p_i}{\sum_{j \in G} p_j e^{\delta} + \sum_{j \in R} p_j}, & \mathcal{V}_i \in \mathcal{V}_R.
\end{cases}
\end{equation}

KGW uses a hyperparameter $\delta$ to control the watermark strength. A larger $\delta$ results in improved watermark effectiveness, but lower text quality. 
By autoregressively sampling from this modified distribution, watermarked sequences can be generated, where the presence of a watermark can be detected based on the proportion of tokens selected from the green list $ \mathcal{V}_G $. 

Specifically for watermark detection, to determine whether a sequence $S = s_1, s_2, \dots, s_{|T|}$ contains a watermark, we calculate the $z$-score as:

\begin{equation}
    z = \frac{|S|_G - \gamma |T|}{\sqrt{|T|\gamma(1-\gamma)}},
\end{equation}

\noindent where $|T|$ is the total number of tokens and $|S|_G$ is the number of tokens in the green list. By setting a threshold of $z$-score, we can determine if the sequence is watermarked. If $z$-score exceeds the threshold, it indicates that the sequence contains a watermark. 

\begin{figure}[t!]
\centering
    \subfigure[$\omega=0.2$]{
        \includegraphics[width=0.2200\textwidth]{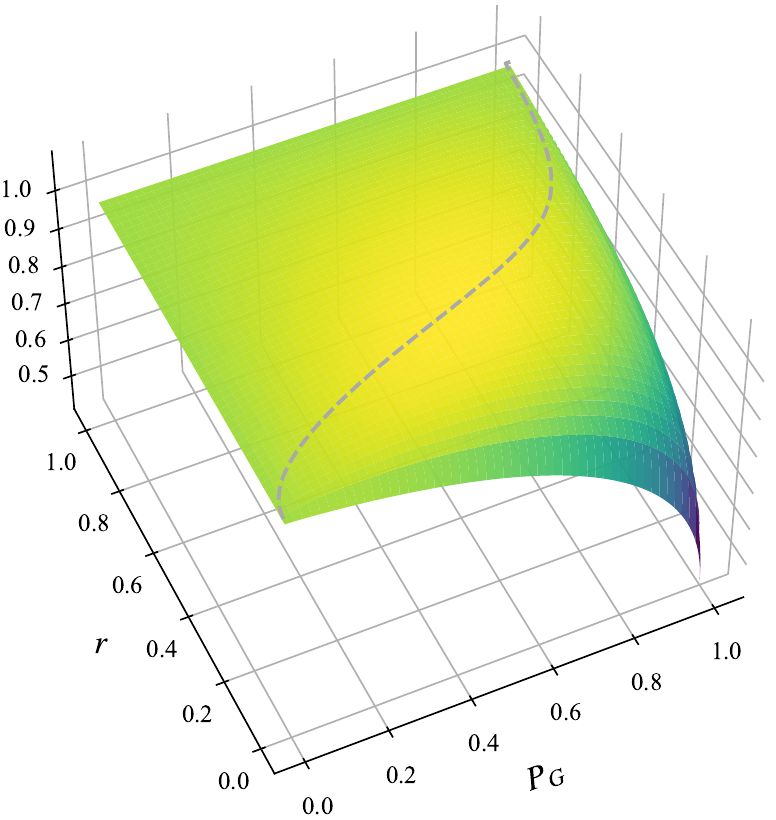}
        \label{fig:Pg_r_F_0_2}
    }
    \subfigure[$\omega=0.4$]{
        \includegraphics[width=0.2200\textwidth]{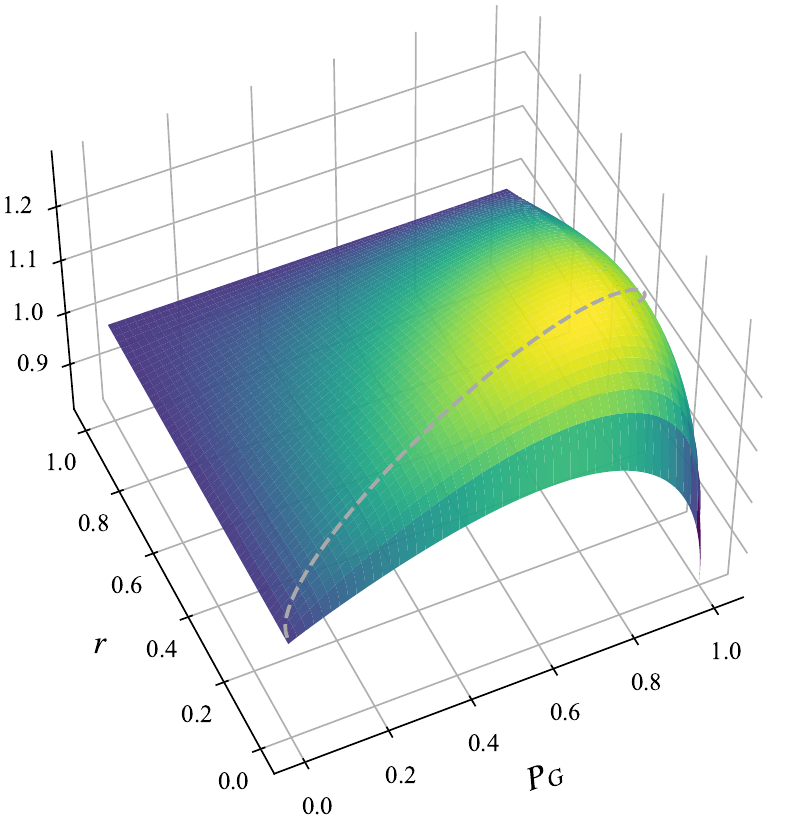}
        \label{fig:Pg_r_F_0_4}
    }
\caption{Visualization of $\mathcal{F}$ across different $P_G$ and $r$. The vertical axis represents $\mathcal{F}$. A dashed dark gray line is used to indicate the optimal $r$ (i.e., $r^*$) that maximizes $\mathcal{F}$ for a fixed $P_G$. We can observe that as $P_G$ decreases, $r^*$ also decreases. }
\label{fig:Pg_r_F}
\end{figure}

\section{Methodology}
\label{sec:methodology}

In this section, we provide a detailed introduction to the proposed watermark method MorphMark. First, in \S ~\ref{sec:trade_off_watermark}, we formalize the multi-objective analysis function $\mathcal{F}$, which can comprehensively capture both text quality $\mathcal{T}$ and watermark effectiveness $\mathcal{W}$. We then theoretically prove that as $P_G$ decreases, reducing $r$ can lead to a larger $\mathcal{F}$. Based on this insight, we describe our watermark algorithm detailedly in \S~\ref{sec:simple_method}.

\subsection{Multi-objective Trade-off Framework}
\label{sec:trade_off_watermark}

In this section, we will model the multi-objective trade-off framework during the proces of sampling the next token. 

\noindent\textbf{Watermark Mechanism.} During generating a new token, we have an original sampling distribution $P = \{{p_i}\}^{|\mathcal{V}|}_{1}$. To watermark this token, we first split the vocabulary into a green list $ \mathcal{V}_G $ and a red list $ \mathcal{V}_R $. Let $ P_G $ represent the sum of probabilities of green tokens, i.e, 

\begin{equation}
P_G = \sum_{j \in G} p_j. 
\end{equation}

Since the maximum increase of $P_G$ is $1 - P_G$ (as $P_G$ cannot exceed 1), we define the total increase of $P_G$ as $r \cdot (1 - P_G)$, where $r$ is used to represent watermark strength and $r \in (0, 1)$. The larger $r$, the greater the watermark strength. Formally, we have the watermarked sampling distribution $\hat{P} = \{\hat{p}_i\}^{|\mathcal{V}|}_{1}$: 

\begin{equation}
\hat{p}_i = 
\begin{cases} 
p_i + \frac{p_i}{P_G} \cdot r (1-P_G), & \mathcal{V}_i \in \mathcal{V}_G, \\
p_i - \frac{p_i}{1-P_G} \cdot r (1-P_G), & \mathcal{V}_i \in \mathcal{V}_R.
\end{cases}
\end{equation}

\noindent\textbf{Text Quality.} Following~\citet{zhaoprovable}, we define text quality as the similarity between original and watermarked sampling distributions. Here we use the Bhattacharyya Coefficient (BC)~\cite{bhattacharyya1946measure, ramesh2023picture} for computational simplicity. Other metrics (e.g., KL divergence) also yield same conclusion, as shown in App.~\ref{sec:proof_r_increase_kl}. 

\begin{equation}
    \begin{split}
\mathcal{T}(r) &= \text{BC}(P, \hat{P}) = \sum_{i \in \mathcal{V}} \sqrt{p_i \hat{p}_i}\\
    = P_G & \sqrt{1+\frac{r\left( 1-P_G \right)}{P_G}}+(1-P_G)\sqrt{1-r},
    \end{split}
\end{equation}

\noindent where \(\mathcal{T}(r)\) represents the BC between \(P\) and \(\hat{P}\). A higher value of \(\mathcal{T}\) indicates a smaller perturbation introduced by the watermark, which corresponds to better preservation of text quality. 

\noindent\textbf{Watermark Effectiveness.} The effectiveness of the watermark can be quantified by the difference between the adjusted probabilities of tokens in the green list and those in the red list. Specifically, it is given by: 

\begin{equation}
\begin{split}
    \mathcal{W}(r) &= (\hat{P}_G - \hat{P}_R) - (P_G - P_R) \\
    &= 2 r (1-P_G),
\end{split}
\end{equation}

\noindent where $\hat{P}_G$ and $\hat{P}_R$ represent the summed probability of tokens in the green and red lists, respectively, under the watermarked sampling distribution, and $P_G$ and $P_R$ correspond to the probabilities under the original sampling distribution. 

\noindent\textbf{Multi-objective Trade-off Analysis Function.} Then, we can construct a multi-objective trade-off analysis function $\mathcal{F}$ as a weighted sum of text quality and watermark effectiveness: 

\begin{equation}
\begin{aligned}
\mathcal{F}(r) &= \mathcal{T}(r)+ \omega \cdot \mathcal{W}(r) \\
  &= P_G \sqrt{1 + \frac{r(1 - P_G)}{P_G}} + (1 - P_G) \sqrt{1 - r} \\
  &\quad + \, \omega \cdot 2 r (1 - P_G),
\end{aligned}
\end{equation}

\noindent where $\omega$ is the weight of watermark effectiveness. We do not impose any restrictions on $\omega$ except $\omega > 0$. Crucially, our subsequent derivations and analysis are valid regardless of the specific value of $\omega$. In other words, whether prioritizing text quality ($\omega$ is small) or watermark effectiveness ($\omega$ is large), our proposed method and conclusions are universally applicable. This can illustrate the wide applicability of our method, enabling it to adapt to various needs and preferences. 

\begin{theorem}\label{thm:r_increase}
Consider the process of sampling a token from the watermarked probability distribution described above, for any given $\omega > 0$, there exists an optimal $r^* \in (0,1)$ that maximizes $\mathcal{F}$. Moreover, the optimal $r^*$ is positively correlated with $P_G$, i.e., $\frac{\partial r^*}{\partial P_G} > 0$.
\end{theorem}

This theorem indicates that, whether prioritizing text quality or watermark effectiveness, adaptively adjusting $r$ in a positively correlated manner with $P_G$ will lead to newly generated tokens achieving both higher text quality and stronger watermark effectiveness. This guides us to adaptively assign larger $r$ when $P_G$ is high, and conversely, smaller $r$ when $P_G$ is low, in order to achieve a larger $\mathcal{F}$. The proof of Theorem~\ref{thm:r_increase} is provided in App.~\ref{sec:proof_r_increase}. 

\noindent\textbf{Visualization of Theoretical Insights.} To provide a straightforward understanding of our insights, we visualize $\mathcal{F}$ in Fig.~\ref{fig:Pg_r_F}. We can clearly observe that no matter how the $\omega$ is set, the lager the $P_G$, the lager the $r$ that maximizes $\mathcal{F}$ (i.e., $r^*$). 

\subsection{Adaptive Watermark}
\label{sec:simple_method}

\begin{figure}[t!]
\centering
    \includegraphics[width=0.482\textwidth]{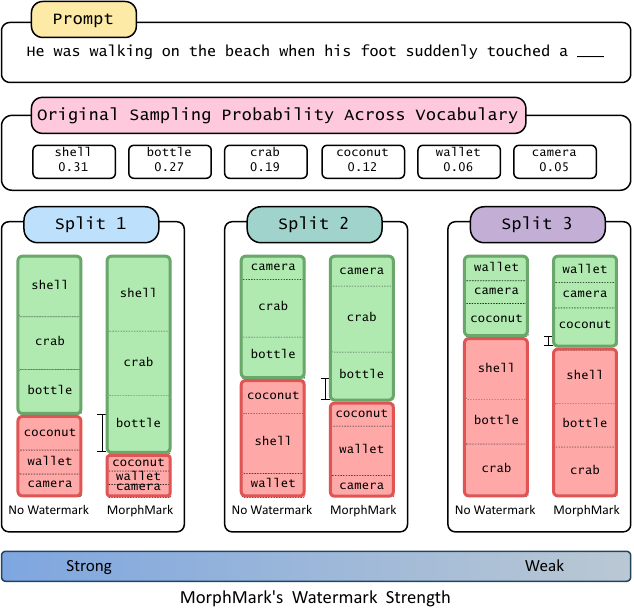}
\caption{An example illustrating the adaptive mechanism of MorphMark. During token generation, the vocabulary is split into green and red lists. Since the split is based on the preceding tokens and user-defined keys, different tokens and users will have different splits. MorphMark adjusts the watermark strength based on the total probability of green tokens. High strength is applied when this probability is high, while low strength is used when this probability is low. }
\label{fig:main_illustration}
\end{figure}

In this section, we propose an instance of the function $ r=\phi(P_G) $ that satisfies the design principle outlined above: 

\begin{equation}
\phi(x) = 
\begin{aligned}
    \begin{cases}
    \epsilon, & x \leq p_0, \\
    \text{min}(z(x), 1-\epsilon), & x > p_0,
\end{cases}
\end{aligned}
\label{eq:linear_phi}
\end{equation}

\begin{equation}
    z(x) = k_{linear}x,
\end{equation}

\noindent where $\epsilon$ is a negligibly small positive value approaching $0$. The function is a piecewise linear function defined over the domain $ (0, 1) $. The parameter $p_0$ is the threshold for watermarking, which we call watermarking threshold. We set $ \phi(P_G) = \epsilon $ when $ x \leq p_0 $, ensuring a very little adjustment to tokens when probabilities in the green list are very small. For $ P_G $ in $(p_0, 1)$, $ \phi(x) $ increases linearly. 
A specific example of this adaptive mechanism used in MorphMark is illustrated in Fig.~\ref{fig:main_illustration}. 

We can also design a fast growth function $ z(x) = e^{k_{exp}x} - 1 $ and a slow growth function $ z(x) = \ln(k_{log}x + 1) $, which we will explore later to determine which approach is better. For detection, we use $z$-score as KGW~\cite{kirchenbauer2023watermark} described in \S~\ref{sec:preliminary}. 
Building on the formula above, we outline the detailed watermark algorithm for text generation in Alg.~\ref{alg:watermark_text_generation} of App.~\ref{sec:algorithm}.

\begin{table*}[t!]
\centering
\renewcommand{\arraystretch}{0.95}
\scalebox{0.772}{
\begin{tabular}{l@{}cccccccc}
\Xhline{1.15pt}
\textbf{Method} & \textbf{TPR@1\%$\uparrow$}      & \textbf{\begin{tabular}[c]{@{}c@{}}TPR@1\%$\uparrow$\\ (Word-S/30\%)\end{tabular}}       & \textbf{Best F1$\uparrow$}         & \textbf{\begin{tabular}[c]{@{}c@{}}Best F1$\uparrow$\\ (Word-S/30\%)\end{tabular}}         & \textbf{PPL$\downarrow$}             & \textbf{\begin{tabular}[c]{@{}c@{}}Generation\\ Time (s)\end{tabular}} & \textbf{\begin{tabular}[c]{@{}c@{}}Detection\\ Time (ms)\end{tabular}} & \textbf{\begin{tabular}[c]{@{}c@{}}Memory\\ Usage (B)\end{tabular}} \\ \Xhline{0.80pt}
\multicolumn{9}{>{\columncolor[gray]{0.9}}c}{\textit{OPT-1.3B}} \\
UnWM  & -               & -               & -               & -               & 10.4815           & 2.4374            & -                 & 0          \\
KGW          & 0.9900 	& 0.8050 	& 0.9950 	& 0.9268 	& 11.4994 
          & 2.4901            & 33.81             & 0          \\
UW           & 1.0000          & 0.7425          & 0.9975          & 0.9221 
          & 11.5854          & 2.5486            & 71.30             & 0          \\
DiPmark      & 0.9975          & 0.7250          & 0.9975          & 0.9138          & 11.5042          & 2.5492            & 71.54             & 0          \\
SWEET        & 0.9975          & 0.8225          & 0.9975          & 0.9501          & 11.5065          & 2.4667            & 44.27             & 1.3        \\
EWD          & 1.0000          & 0.8450          & 1.0000          & 0.9549          & 11.4777          & 2.4526            & 44.52             & 1.3        \\
MorphMark$_{exp}$	& \textbf{1.0000} 	& \textbf{0.9600} 	& 0.9975 	& \textbf{0.9778} 	& 11.3569 	& 2.6768 	& 34.17 	& 0 \\
MorphMark$_{linear}$	& \textbf{1.0000} 	& 0.9275 	& 0.9962 	& 0.9727 	& \textbf{11.2386} 	& 2.6537 	& 33.99 	& 0 \\
MorphMark$_{log}$	& \textbf{1.0000} 	& 0.9375 	& \textbf{1.0000} 	& 0.9660 	& 11.3379 	& 2.6889 	& 34.45 	& 0 \\ \Xhline{0.80pt}
\multicolumn{9}{>{\columncolor[gray]{0.9}}c}{\textit{OPT-2.7B}} \\
UnWM  & -               & -               & -               & -               & 9.6683           & 3.1573            & -                 & 0          \\
KGW          & 0.9950          & 0.8275          & 0.9950          & 0.9098          & 10.9324          & 3.2353            & 33.01             & 0          \\
UW           & 0.9950          & 0.6900         & 0.9962          & 0.9202          & 10.8593          & 3.3178            & 72.86             & 0          \\
DiPmark      & 0.9900          & 0.7125          & 0.9913          & 0.9058          & 11.0013          & 3.3126            & 72.83             & 0          \\
SWEET        & 0.9975          & 0.8350          & 0.9962          & 0.9566          & 10.8377          & 3.2605            & 49.46             & 2.7        \\
EWD          & 1.0000          & 0.8500          & 0.9988 & 0.9588          & 10.6303          & 3.2180            & 49.56             & 2.7        \\
MorphMark$_{exp}$	& \textbf{1.0000} 	& \textbf{0.9625} 	& 0.9987 	& 0.9686 	& 10.5144 	& 3.5074   & 34.64 	& 0  \\
MorphMark$_{linear}$	& \textbf{1.0000} 	& 0.9300 	& \textbf{0.9988} 	& \textbf{0.9701} 	& \textbf{10.3852} 	& 3.4149 	& 34.00 	& 0  \\
MorphMark$_{log}$	& 0.9975 	& 0.9250 	& \textbf{0.9988} 	& 0.9628 	& 10.6717 	& 3.6792 	& 34.63 	& 0 \\ \Xhline{0.80pt}
\multicolumn{9}{>{\columncolor[gray]{0.9}}c}{\textit{OPT-6.7B}} \\
UnWM  & -               & -               & -               & -               & 9.0120           & 4.2656            & -                 & 0          \\
KGW          & 0.9950          & 0.8150          & 0.9975          & 0.9058          & 9.9602           & 4.3163            & 32.30             & 0          \\
UW           & 0.9950          & 0.7025          & 0.9899          & 0.8971          & 10.3701          & 4.4407            & 75.04             & 0          \\
DiPmark      & 0.9975          & 0.6625          & 0.9925          & 0.9073          & 10.2747          & 4.4363            & 75.13             & 0          \\
SWEET        & 0.9925          & 0.7925          & 0.9975          & 0.9539          & 10.0633          & 4.3931            & 62.20             & 6.7        \\
EWD          & 1.0000          & 0.8350          & 0.9975          & 0.9523          & 9.9925           & 4.3393            & 61.74             & 6.7        \\
MorphMark$_{exp}$	& \textbf{1.0000} 	& 0.9100 	& \textbf{0.9975} 	& \textbf{0.9763} 	& \textbf{9.6618} 	& 4.5198 	& 35.97 	& 0  \\
MorphMark$_{linear}$	& 0.9975 	& \textbf{0.9250} 	& 0.9950 	& 0.9637 	& 9.7391 	& 4.4456 	& 35.15 	& 0  \\
MorphMark$_{log}$	& 0.9950 	& 0.8975 	& 0.9950 	& 0.9602 	& 9.8585 	& 4.4537 	& 35.45   & 0 \\
\Xhline{1.15pt}
\end{tabular}}
\caption{Performance comparison on different methods. The best results are in bold for each column. }
\label{tab:overall_performance}
\end{table*}

\section{Experiments}

\subsection{Experimental Setup}

Following MarkLLM~\cite{pan2024markllm}, we evaluate MorphMark using 400 samples from the C4~\cite{raffel2020exploring}, with OPT-1.3B, -2.7B, and -6.7B~\cite{zhang2022opt} as the backbone models. Our baselines include various flexible watermark methods, including KGW~\cite{kirchenbauer2023watermark}, UW~\cite{hu2024unbiased}, DiPmark~\cite{wu2024resilient}, SWEET~\cite{lee2024wrote}, and EWD~\cite{lu2024entropy}. We assess watermark effectiveness in terms of detectability (TPR@1\%, Best F1) and robustness (assessed under the Word-S/30\% attack, where 30\% of words are randomly replaced with synonyms from WordNet), as well as text quality via perplexity (PPL). Details are shown in App~\ref{sec:additional_experimental_setup}.

\subsection{Overall Performance}

We summary the main results in Tab.~\ref{tab:overall_performance}. Besides watermark effectiveness and text quality, we report the time spent on generation~(Generation Time~(s)) and detection~(Detection Time~(ms)) (for 800 tokens), as well as the size of models used for detection~(Memory Usage~(B)) to highlight the time and space efficiency of different watermark methods. 

From the results, we can see that MorphMark outperforms all baselines in detectability, robustness, and text quality, demonstrating a superior effectiveness-quality trade-off.
It spends nearly identical generation and detection time to that of KGW, indicating no significant additional delay. Additionally, MorphMark incurs no memory usage during detection, as it does not require loading any model. In summary, MorphMark is an efficient method that effectively address the dilemma between watermark effectiveness and text quality.

\subsection{Performance on Robustness}

\begin{figure*}[t!]
\centering
    \includegraphics[width=1.0\textwidth]{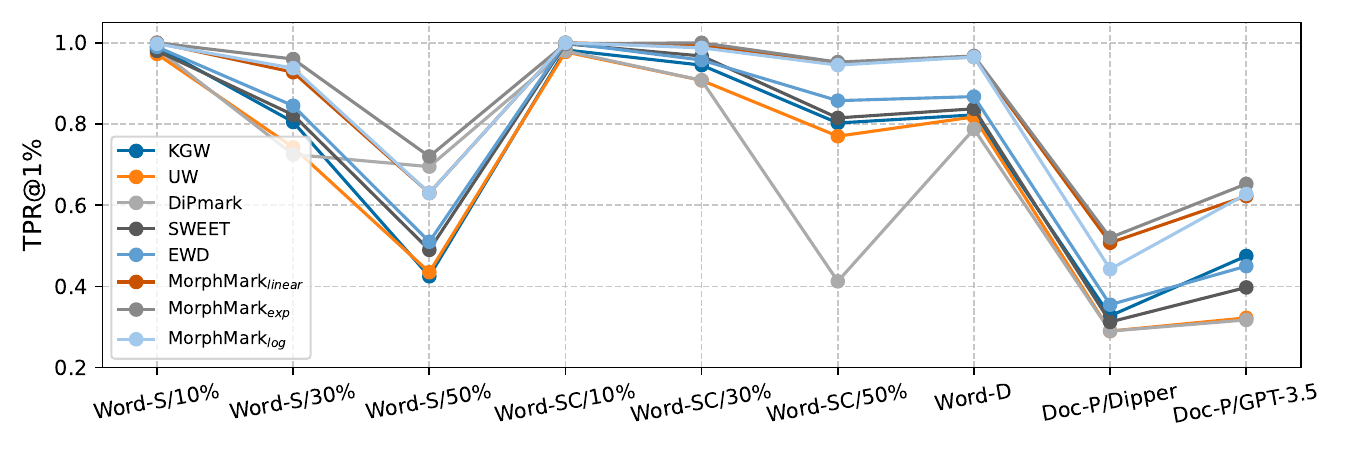}
\caption{Robustness performance of each watermarking method under various attack scenarios. }
\label{fig:robustness_performance}
\end{figure*}
Malicious attackers may use paraphrasing attack methods to conduct watermark removal. Thus, we implement 5 paraphrasing attack methods to evaluate the robustness of different watermarking algorithms.
(1) Word-S/ refers to randomly replacing words with synonyms from WordNet, where the number after "/" indicates the proportion of words modified. (2) Word-SC/ refers to randomly replacing words with synonyms from WordNet based on context. (3) Word-D involves randomly deleting 30\% of the words from the text. (4) Doc-P (GPT-3.5) rewrites the text using GPT-3.5-Turbo~\cite{openai_gpt35}. Details are shown in App.~\ref{sec:gpt35}. (5) Doc-P (Dipper) rewrites the text using a specialized paraphrasing model Dipper~\cite{krishna2024paraphrasing}. 

We summarize the results in Fig.~\ref{tab:overall_performance}. As shown, MorphMark$_{exp}$ exhibits significantly superior robustness compared to all other methods across all attack scenarios. This advantage is particularly evident when watermarked texts are paraphrased by GPT-3.5 or Dipper, where MorphMark$_{exp}$ achieves a substantially higher TPR@1\%. In addition, the other two variants, MorphMark$_{linear}$ and MorphMark$_{log}$, also outperform the selected baselines in most attack settings. 
In summary, these results empirically demonstrate the strong robustness of MorphMark, particularly MorphMark$_{exp}$, making it a more practical and reliable choice. 

\subsection{Performance on Text Quality}

Following previous work~\cite{hu2024unbiased, wu2024resilient}, instead of using PPL only, we evaluate text quality on two downstream tasks, specifically machine translation and text summarization. For machine translation, we employ the nllb-200-distilled-600M~\cite{costa2022no} as our translation model and randomly sample 400 instances from the WMT16~\cite{bojar2016findings} corpus for the German-to-English translation task as our test dataset. For text summarization, we evaluate 400 randomly sampled instances from the CNN-DM dataset~\cite{hermann2015teaching} using the OPT-1.3B model~\cite{zhang2022opt}. To assess performance, we employ BLEU~\cite{papineni2002bleu}, ROUGE~\cite{lin2004rouge}, and BERTScore~\cite{zhang2019bertscore} as evaluation metrics. Our experiments use the same parameters as the main study, ensuring that text quality is compared under the condition that MorphMark's detectability and robustness surpass other watermarking methods. 

\begin{figure}[h!]
\centering
    \subfigure[Machine Translation]{
        \hspace{-1mm}
        \includegraphics[width=0.4900\textwidth]{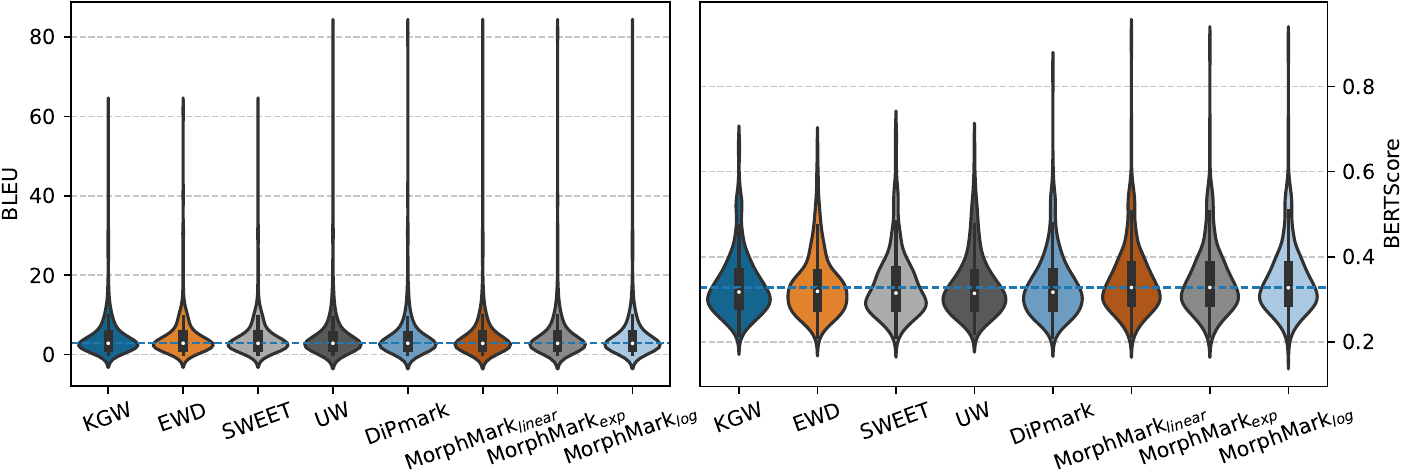}
        \label{fig:text_quality_MT}
    }\\
    \subfigure[Text Summarization]{
        \hspace{-1mm}
        \includegraphics[width=0.4900\textwidth]{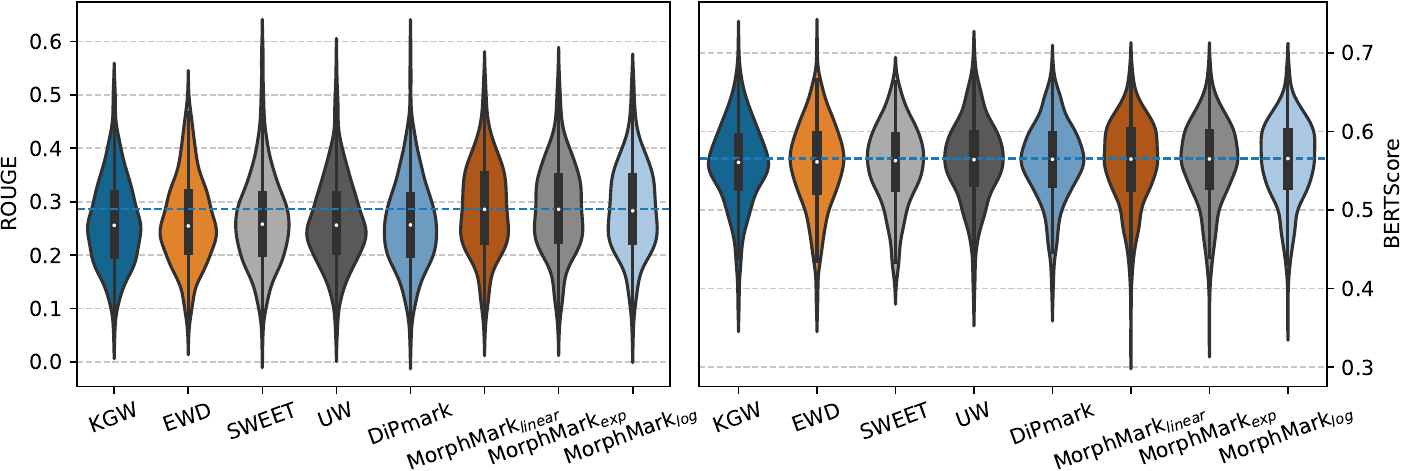}
        \label{fig:text_quality_TS}
    }
\caption{Text quality on downstream tasks. }
\label{fig:text_quality}
\end{figure}

\begin{figure*}[t!]
\centering
    \subfigure[$k_{linear}$ (MorphMark$_{linear}$)]{
        \includegraphics[width=0.2455\textwidth]{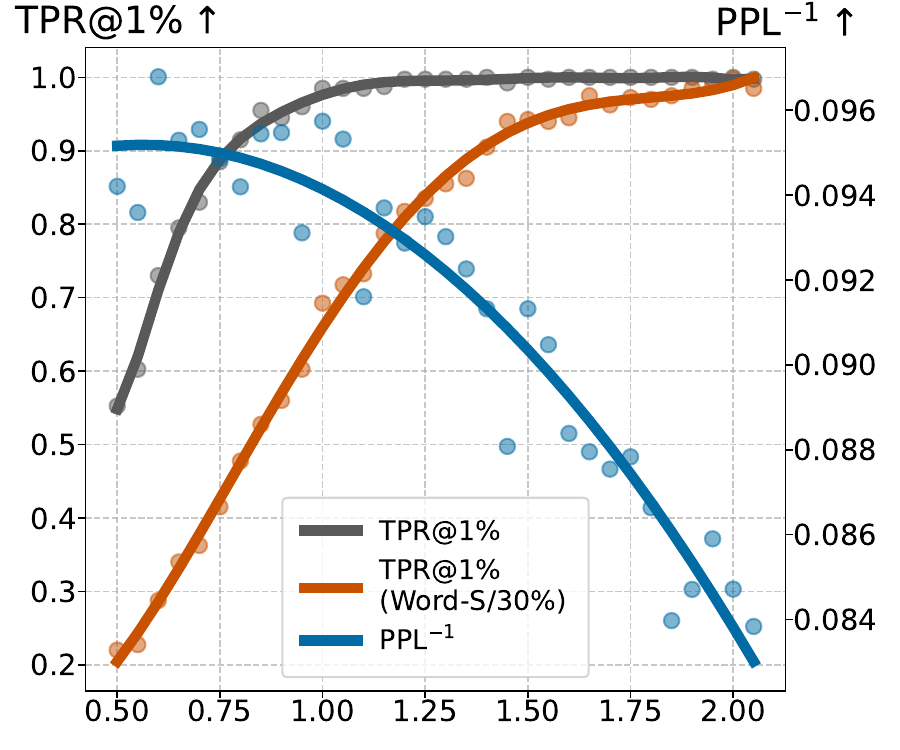}
        \label{fig:ablation_1}
    }\hspace{-2.5mm}
    \subfigure[$k_{exp}$ (MorphMark$_{exp}$)]{
        \includegraphics[width=0.2455\textwidth]{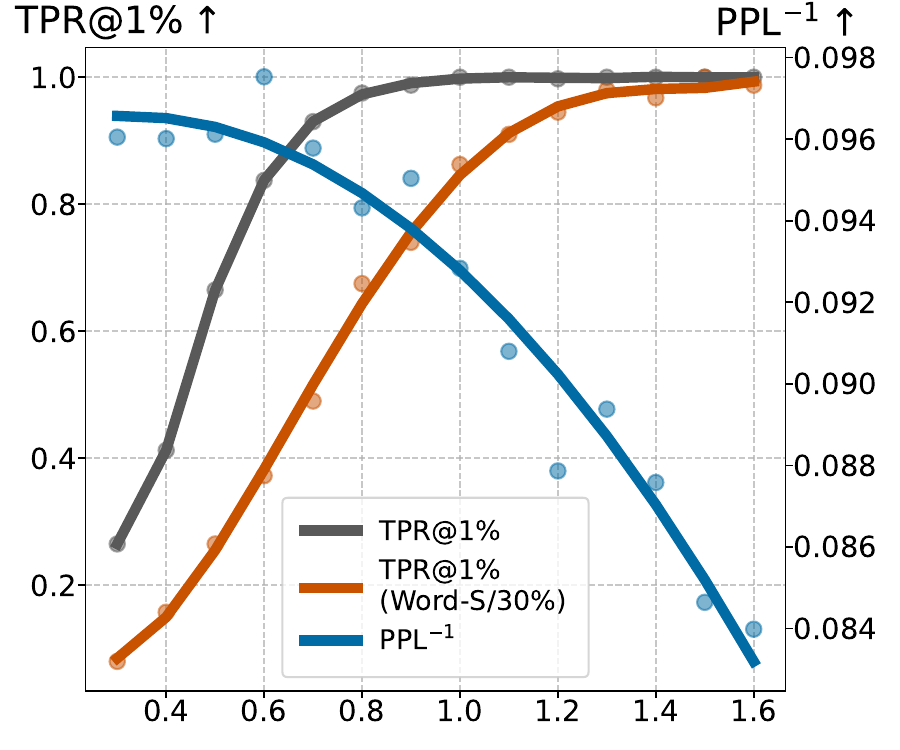}
        \label{fig:ablation_2}
    }\hspace{-2.5mm}
    \subfigure[$k_{log}$ (MorphMark$_{log}$)]{
        \includegraphics[width=0.2455\textwidth]{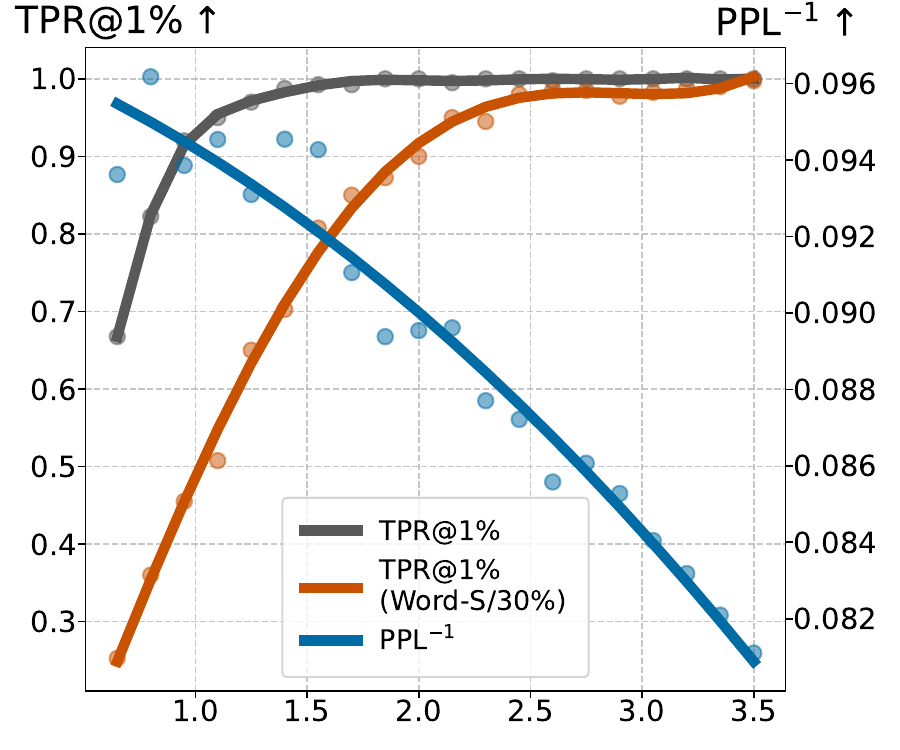}
        \label{fig:ablation_3}
    }\hspace{-2.5mm}
    \subfigure[$p_{0}$ (MorphMark$_{exp}$)]{
        \includegraphics[width=0.2455\textwidth]{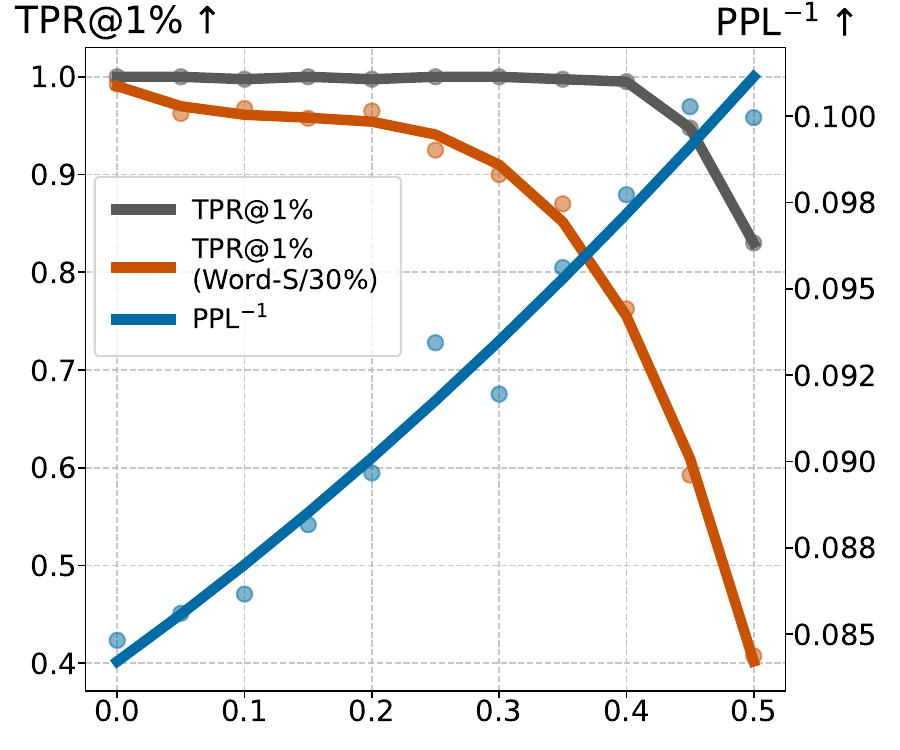}
        \label{fig:ablation_4}
    }
\caption{\textbf{Parameter ablation study of MorphMark.} In (a), (b), and (c), we conduct an ablation study on $k$ across different variants of MorphMark, where the x-axis represents $k$. In (d), we perform an ablation study on the watermarking threshold, where the x-axis represents $p_0$.}
\label{fig:ablation}
\end{figure*}

Fig.~\ref{fig:text_quality_MT} presents the results for the machine translation task. In terms of the BLEU metric, all methods demonstrate comparable performance. However, for BERTScore, our proposed method, MorphMark's three variants, consistently outperforms all other baseline methods by a small margin. 
Fig.~\ref{fig:text_quality_TS} shows the results for the text summarization task. According to the ROUGE metric, the MorphMark$_{exp}$ and MorphMark$_{linear}$ variants exhibit slightly better performance than the MorphMark$_{log}$ variant, while all three significantly outperform the baseline methods. For BERTScore, the three MorphMark variants yield nearly identical performance, showing a minor improvement over the unbiased watermarks (UW and DiPmark). Furthermore, both MorphMark and unbiased watermarks achieve a notable advantage over the other baseline approaches. 

Overall, in terms of text quality, MorphMark outperforms unbiased watermarks (UW and DiPmark), and these two unbiased watermarks surpasses all other baseline approaches.

\subsection{Ablation Study}
In this section, we conduct ablation study on the hyper-parameters of MorphMark, including $ k_{exp} $, $ k_{linear} $ and $ k_{log} $ in MorphMark$_{exp}$, MorphMark$_{linear}$, and MorphMark$_{log}$ respectively, as well as $ p_0 $.
The impact of these parameters is clearly shown in Fig.~\ref{fig:ablation}. Specifically, as $ k_{exp} $, $ k_{linear} $ and $ k_{log} $ increase, or as $ p_0 $ decreases, watermark strength increase, so watermark effectiveness improve, while text quality degrades. 

Additionally, by combinating Fig.\ref{fig:ablation_1}, Fig.\ref{fig:ablation_2}, and Fig.~\ref{fig:ablation_3}, we can conveniently compare MorphMark$_{exp}$, MorphMark$_{linear}$, and MorphMark$_{log}$. By fixing either watermark effectiveness or text quality, we can assess the relative performance of the three variants along the other dimension. This analysis leads to the conclusion that across various levels of detectability, the text quality ranking consistently follows MorphMark$_{exp}$ > MorphMark$_{linear}$ > MorphMark$_{log}$. This highlights MorphMark$_{exp}$'s superior trade-off between watermark effectiveness and text quality, making it the strongest choice among the three designs. 

\subsection{Further Analyses}

\subsubsection{Different Sampling Parameters}

In this section, we test whether MorphMark remains effective under different sampling parameters. We consider several commonly used temperature and top-p combinations: (1.2, 1.0) for high creativity, (0.7, 0.95) and (0.9, 0.95) for general-purpose tasks, and (0.3, 1.0) for precision-oriented tasks. 

\begin{table}[h!]
\centering
\scalebox{0.63}{
\begin{tabular}{ccccc}
\Xhline{1.15pt}
\textbf{\begin{tabular}[c]{@{}c@{}}(Temp, TopP)\end{tabular}} & \textbf{UnWM PPL} & \textbf{PPL}     & \textbf{TPR@1\%} & \textbf{\begin{tabular}[c]{@{}c@{}}TPR@1\%$\uparrow$\\ (Word-S/30\%)\end{tabular}} \\ \hline
(0.3, 1.0)         & 4.1308   & 4.7605  & 0.9925  & 0.9200  \\
(0.7, 0.95)        & 5.4809   & 6.1871  & 1.0000  & 0.9450  \\
(0.9, 0.95)        & 7.3829   & 8.0190  & 0.9975  & 0.9550  \\
(1.2, 1.0)         & 15.2175  & 16.8605 & 0.9975  & 0.9600 \\
\Xhline{1.15pt}
\end{tabular}}
\caption{Performance of MorphMark$_{exp}$ with different sampling parameters. UnWM refers to unwatermarked output. }
\label{tab:sampling_param_exp}
\end{table}

Table~\ref{tab:sampling_param_exp} presents the results of MorphMark$_{exp}$. From the results, we observe that as the temperature increases, both the unwatermarked PPL and watermarked PPL increase, indicating that higher temperature leads to more diverse generations. Additionally, the TPR@1\% remains consistently high across all settings, demonstrating the robustness of MorphMark$_{exp}$. Notably, the relative improvement in TPR@1\% increases with temperature, with the highest improvement observed at (1.2, 1.0), suggesting that watermark detection benefits from more diverse text generation. 

These results indicate that MorphMark$_{exp}$ performs still reliably across different sampling settings, maintaining high detection effectiveness while adapting to different decoding parameters. Results of MorphMark$_{linear}$ and MorphMark$_{log}$ are present in Tab.~\ref{tab:sampling_param_linear} and Tab.~\ref{tab:sampling_param_log} of App.~\ref{sec:sampling_param}. 

\begin{figure*}[t!]
\centering
    \subfigure[Story creation task with widely balanced $P_G$ values. ]{
        \includegraphics[width=0.97\textwidth]{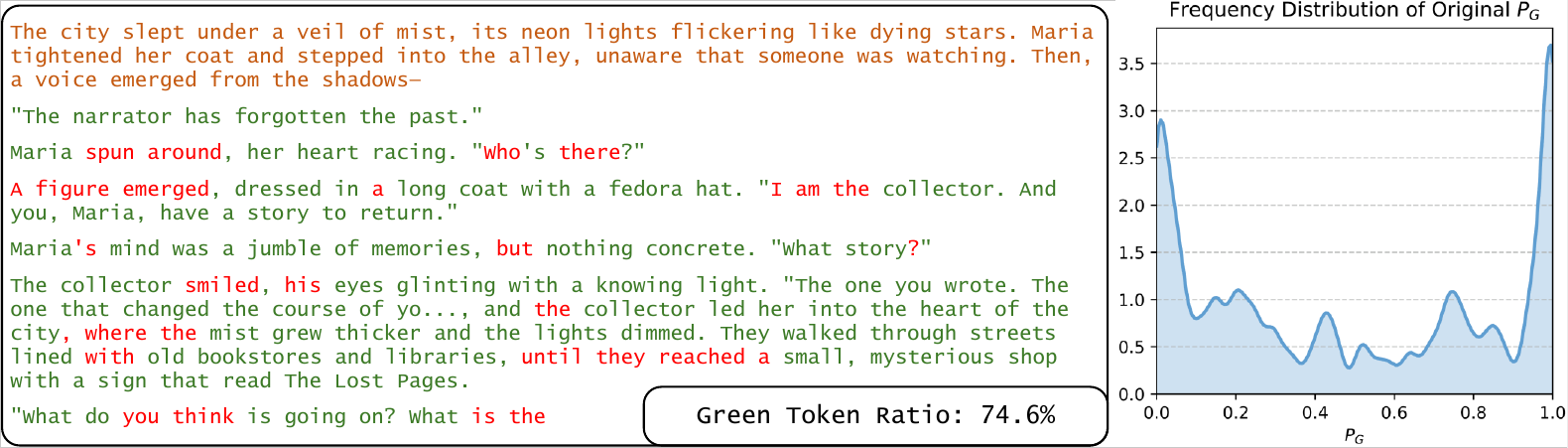}
        \label{fig:case_study_1}
    }
    \subfigure[Code generation task with a high number of extreme $P_G$ values, most $P_G$ values being concentrated near 0 or 1. ]{
        \includegraphics[width=0.97\textwidth]{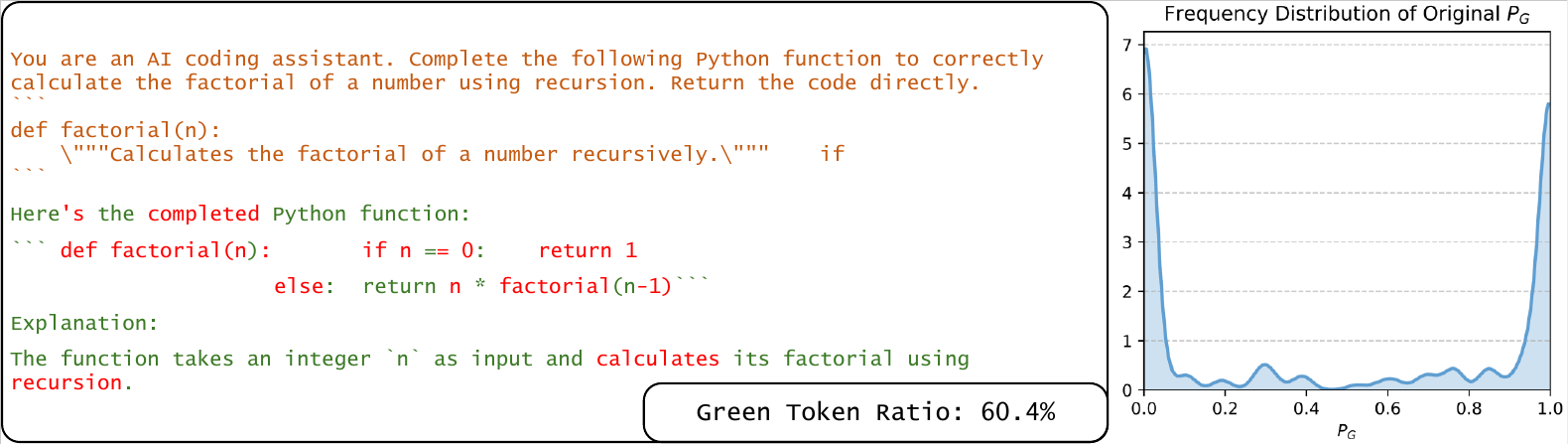}
        \label{fig:case_study_2}
    }
\caption{Case Study on \( P_G \) Distributions. In example (a), which illustrates a story creation task, the \( P_G \) values are well-balanced across a wide range. MorphMark performs effectively in this scenario, achieving a high ratio of green tokens throughout the sequence. In contrast, example (b) presents a code generation task with an extreme distribution, where most \( P_G \) values are concentrated near 0 or 1. In this case, MorphMark proves less effective. }
\label{fig:case_study}
\end{figure*}

\subsubsection{In-Depth Analysis of $P_G$ Distribution}

An important factor affecting the performance of MorphMark is the distribution of $P_G$ within a sequence. For example, if the sequence's entropy is low, $P_G$ tends to concentrate around 0 and 1, making it difficult for MorphMark to successfully inject the watermark. 

Since the distribution of $P_G$ within a sequence is difficult to quantify with a single metric, we present a case study in this section to shed light on this aspect. To this end, we employ two contrasting examples: a high-entropy task, specifically story creation, and a low-entropy task, code generation in Fig.~\ref{fig:case_study}. From these examples, we observe that when the distribution of $P_G$ is extreme, the effectiveness of the watermark is low. 

To determine whether such extreme conditions occur frequently, we examine the distribution of $P_G$ across several popular benchmarks including TruthfulQA~\cite{lin2021truthfulqa}, SQuAD~\cite{rajpurkar2016squad, rajpurkar2018know}, GSM8K~\cite{cobbe2021gsm8k} and MBPP~\cite{austin2021program}. The statistical results are presented in Fig.~\ref{fig:PG_dis}. These results show that the $P_G$'s distribution in most benchmarks is relatively uniform—even in code tasks. This uniformity is likely due to the fact that code typically contains comments, and after alignment, LLMs tend to output additional natural language explanations rather than only code. Overall, since such extreme cases occur infrequently, our method remains effective in most scenarios. 


\subsection{Evaluation in Low-Entropy Scenarios}

To validate the performance of MorphMark in low-entropy scenarios, we conduct experiments on HumanEval~\cite{chen2021evaluating}. We employ StarCoder2-3B~\cite{lozhkov2024starcoder}, a widely used code generation model, as the test model. In this evaluation, we compare three methods: SWEET, EWD and MorphMark$_{exp}$. Given that MorphMark and EWD operate at different stages, we also explore a combined approach, where MorphMark is used for watermark generation, while EWD is employed for watermark detection. We use pass@1 as the evaluation metric. 

\begin{table}[t]
    \centering
    \resizebox{0.45\textwidth}{!}{
    \begin{tabular}{lccc}
        \toprule
        Method & TPR@1$\uparrow$ & Best F1$\uparrow$ & pass@1$\uparrow$ \\
        \midrule
        No Watermark        & -      & -      & 0.2500 \\
        SWEET               & 0.8110 & 0.8926 & 0.1890 \\
        EWD                 & 0.8537 & 0.9180 & 0.2073 \\
        MorphMark           & 0.8780 & 0.9320 & 0.2378 \\
        MorphMark + EWD     & \textbf{0.9207} & \textbf{0.9557} & 0.2378 \\
        \bottomrule
    \end{tabular}
    }
    \caption{Performance comparison of different methods in low-entropy scenario on HumanEval.}
    \label{tab:low_entropy_results}
\end{table}

The results of the experiments are summarized in Table~\ref{tab:low_entropy_results}. The results indicate that MorphMark consistently outperforms SWEET and EWD in low-entropy scenarios. Notably, the combination of MorphMark and EWD achieves the best overall performance, demonstrating that these methods can mutually enhance each other. This finding motivates future research to design adaptive generation and detection methods simultaneously. 
In addition, we conduct an analytical experiment targeting low-entropy scenarios to investigate the mechanisms of SWEET, EWD, and MorphMark, as detailed in App~\ref{sec:analysis_low_entropy}. 

\section{Related Work}

Backdoor-based watermarking has been widely studied before the rise of large language models~\cite{adi2018turning, li2022untargeted, wang2024espew}. 
In the era of LLMs, due to the high cost of training models, researchers have shifted to injecting watermarks during the generation process~\cite{kirchenbauer2023watermark, kuditipudi2024robust}. Recent studies focus on low-entropy watermarking~\cite{lu2024entropy, mao2024watermark}, watermark security~\cite{pang2024no, liu2024sir, he2024xsir}, watermark privacy~\cite{jovanovicwatermark, christ2024undetectable}, and watermark under different sampling methods~\cite{hu2024inevitable, dathathri2024scalable},  with the most widely explored topic being the trade-off between watermark effectiveness and text quality~\cite{hu2024unbiased, wu2024resilient, huo2024token}. 
The full related work is shown in App.~\ref{sec:full_related_work}. 


\section{Conclusion}

This work investigates the fundamental trade-off between watermark effectiveness and text quality when watermarking large language models (LLMs). We first formally characterize this trade-off as a multi-objective analysis function and identify the cumulative probability of green-list tokens as a critical factor influencing this trade-off. Our theoretical analysis reveals that increasing watermark strength does not always lead to improved performance, particularly when the cumulative probability of the green list is low. 
Motivated by this theoretical insight, we introduce MorphMark, a dynamic watermarking mechanism that adaptively adjusts watermark strength to improve both watermark effectiveness and text quality. In addition, MorphMark offers flexibility and efficiency (time and space). Empirical results demonstrate MorphMark's substantial improvement across diverse models and scenarios. 
By integrating theoretical modeling, algorithmic design and innovation, empirical validation, and practical deployment consideration, this work propose a reliable and practical watermarking mechanism. Our findings deepen the understanding of watermarking mechanism based on green-red list and provide the community with both theoretical analytical tool and practical methodology.

\section*{Limitations}

While our empirical analysis demonstrates that MorphMark is effective in a wide range of scenarios, it is important to acknowledge certain limitations. One notable constraint arises in extremely low-entropy text generation tasks, where the watermarking capability of MorphMark becomes nearly less effective. This issue is not unique to MorphMark but rather a fundamental limitation shared by all green-red list-based watermarking methods. 
The core reason behind this limitation lies in the nature of low-entropy text generation. When a model produces highly predictable sequences with minimal variation, the opportunities for embedding watermarks become significantly reduced. Since green-red list-based watermarking relies on a degree of token unpredictability to manipulate token selection probabilities, it struggles to function effectively when entropy is too low. 

Addressing this challenge requires exploring alternative watermarking strategies that do not depend solely on token-level entropy. Potential directions include integrating semantic or syntactic watermarking techniques, leveraging sentence-level perturbations, or incorporating watermark signals at deeper structural levels within the model. 

Despite this limitation, MorphMark remains highly effective in most practical applications. The broad distribution of $P_G$ observed in our experiments suggests that, under typical generation conditions, MorphMark consistently embeds reliable watermarks. Future work should focus on refining watermarking methods to enhance performance in extreme cases while maintaining MorphMark’s efficiency and usability across diverse text generation tasks. 



\bibliography{custom}

\newpage
\onecolumn
\appendix

\section{Algorithm}
\label{sec:algorithm}

We present the detailed algorithm in Alg.~\ref{alg:watermark_text_generation}. 

\begin{algorithm}[!h]
    \caption{Text Generation with Watermark}
    \label{alg:watermark_text_generation}
    \begin{algorithmic}[1]
        \STATE \textbf{Input:} prompt $s_{-N_p}, \dots, s_{-1}$, a private key $k$, hyper-parameters used in Equation~\ref{eq:linear_phi}: $p_{0}$, $k_{linear}$ and $\epsilon$.
        \STATE \textbf{Output:} watermarked text.
        \FOR{$t = 0$ \TO $T$}
            \STATE Obtain the probability distribution vector $p = P\left( s_t \mid \boldsymbol{s}_{-N_{p}:t-1} \right)$ from the language model.
            \STATE Compute a hash value of token $\boldsymbol{s}_{t-1}$ using the private key $k$.
            \STATE Randomly partition the vocabulary into a green list $\mathcal{V}_G$ of size $|\mathcal{V}|/2$ and a red list $\mathcal{V}_R$ of size $|\mathcal{V}|/2$, with the hash value serving as the random seed.
            \STATE Calculate total adjustment $r=\phi(\sum_{j \in G} p_j)$ as defined in Equation~\ref{eq:linear_phi}.
            \STATE Generate the watermarked probability distribution over the vocabulary:
            $$
            \hat{p}_i = 
            \begin{cases} 
            p_i + \frac{p_i}{\sum_{j \in G} p_j} \cdot r \sum_{j \in R} p_j, &\mathcal{V}_i \in \mathcal{V}_G, \\
            p_i - \frac{p_i}{\sum_{j \in R} p_j} \cdot r \sum_{j \in R} p_j, &\mathcal{V}_i \in \mathcal{V}_R.
            \end{cases}
            $$
            \STATE Sample the next token $s_{t}$ based on the watermarked distribution $\hat{p}$.
        \ENDFOR
        \RETURN $\boldsymbol{s}_{0:T}$. 
    \end{algorithmic}
\end{algorithm}

\section{Proof}
\label{sec:proof}

\subsection{Proof of Theorem~\ref{thm:r_increase}}
\label{sec:proof_r_increase}

For simplicity in calculation, we define text quality as the Bhattacharyya coefficient coefficient (BC) between the original sampling distribution and the watermark sampling distribution. Note that using KL divergence also leads to the same conclusion, based on the same derivation process. 

\begin{equation}
    \begin{split}
    \mathcal{T}(r)&=\text{BC}(P, \hat{P}) = \sum_{i \in \mathcal{V}} \sqrt{p_i \hat{p}_i}\\
    &= \sum_{i\in G}{\sqrt{p_i\left( p_i+\frac{p_i}{P_G}r\left( 1-P_G \right) \right)}}+\sum_{i\in R}{\sqrt{p_i\left( p_i-\frac{p_i}{1-P_G}r\left( 1-P_G \right) \right)}}\\
    &= \sum_{i\in G}{p_i\sqrt{1+\frac{r\left( 1-P_G \right)}{P_G}}}+\sum_{i\in R}{p_i\sqrt{1-r}}\\
    &= P_G\sqrt{1+\frac{r\left( 1-P_G \right)}{P_G}}+(1-P_G)\sqrt{1-r}
    \end{split}
\end{equation}

Detection capability is defined as the difference of increased probability of green list and red list: 

\begin{equation}
\mathcal{W}(r) = 2\omega r (1-P_G)
\end{equation}

Thus, we define the multi-objective trade-off analysis function as a weighted sum of both: 

\begin{equation}
\mathcal{F} = \mathcal{T} + \omega \cdot \mathcal{W} = P_G\sqrt{1+\frac{r\left( 1-P_G \right)}{P_G}}+(1-P_G)\sqrt{1-r} + 2\omega r (1-P_G)
\end{equation}

where $\omega$ is the weight of detection capability and $\omega > 0$. For generality, we impose no additional restrictions on $\omega$. That is, our following derivation is valid for any $w$. 

The first derivative of $\mathcal{F}$ with respect to $r$ is:

\begin{equation}
\frac{\partial \mathcal{F}}{\partial r} = (1-P_G)\left( 2\omega +\frac{1}{2\sqrt{1+\frac{r \left( 1-P_G \right)}{P_G}}}-\frac{1}{2\sqrt{1-r}} \right) 
\end{equation}

We only need the sign of the derivative later. To simplify the calculation, we use $S$ to replace the derivative above, as $S$ has the same sign. 

\begin{equation}
S = 2\omega +\frac{1}{2\sqrt{1+\frac{r \left( 1-P_G \right)}{P_G}}}-\frac{1}{2\sqrt{1-r}}
\end{equation}

Next, we need to prove that $\mathcal{F}$ achieves its maximum at $S = 0$. The formula for the first derivative of $S$ with respect to $r$ is: 

\begin{equation}
    \begin{split}
    \frac{\partial S}{\partial r} &= \frac{1}{4\cdot \left( -r +1+\frac{r}{P_G} \right) ^{\frac{3}{2}}}-\frac{1}{4\cdot \left( 1-r \right) ^{\frac{3}{2}}}-\frac{1}{4\cdot P_G\cdot \left( -r +1+\frac{r}{P_G} \right) ^{\frac{3}{2}}}\\
    &= -\frac{1-P_G}{4P_G\left( 1+r \left( \frac{1}{P_G}-1 \right) \right) ^{\frac{3}{2}}}-\frac{1}{4\left( 1-r \right) ^{\frac{3}{2}}} < 0
    \end{split}
\end{equation}

This derivative is negative, meaning that $S$ is decreasing as $r$ increases. 


\begin{equation}
\lim_{r \to 0} S = 2 \omega > 0
\end{equation}

\begin{equation}
\lim_{r \to 1} S = -\infty < 0
\end{equation}

Since $S$ is positive at $r = 0$ and negative at $r = 1$, by the continuity of $S$ and the intermediate value theorem, there must exist a value $r^*$ between 0 and 1 such that $S(r^*) = 0$. 

By the implicit function theorem, substituting $S = 0$, we obtain the relationship between $r^*$ and $P_G$:

\begin{equation}
\frac{\partial r^{*}}{\partial P_G} =  -\frac{\frac{\partial S}{\partial P_G}}{\frac{\partial S}{\partial r^{*}}} = -\frac{\frac{r^{*}}{4\cdot P_{G}^{2}\cdot \left( 1+r^{*}\left( \frac{1}{P_G}-1 \right) \right) ^{\frac{3}{2}}}}{-\frac{1-P_G}{4P_G\left( 1+r^{*}\left( \frac{1}{P_G}-1 \right) \right) ^{\frac{3}{2}}}-\frac{1}{4\left( 1-r^{*} \right) ^{\frac{3}{2}}}} > 0
\end{equation}

This shows that when $P_G$ is larger, the optimal $r$ will also be larger, and when $P_G$ is smaller, the optimal $r$ will be smaller. In other words, increasing $P_G$ leads to an increase in the optimal parameter $r^*$. This implies that in the optimization process, as the sampling distribution changes, the watermark optimization parameter $r$ needs to be adjusted accordingly to maintain optimal performance. 

\subsection{Proof for More Similarity Measurement Methods}
\label{sec:proof_r_increase_kl}

Here, we use another similarity measurement method (KL divergence) to measure the text quality. And we will prove that it also leads to the same conclusion. Since we need the similarity instead of divergence, so we calculate $-D_{KL}(P || \hat{P})$: 

\begin{equation}
    \begin{split}
    \mathcal{T}(r)&=-D_{KL}(P || \hat{P}) = \sum_{i \in \mathcal{V}}p_{i}log\frac{\hat{p}_i}{p_i} \\
    &= \sum_{i \in G}p_{i}log\frac{p_{i}+\frac{p_{i}}{P_G}r(1-P_G)}{p_i} + \sum_{i \in R}p_{i}log\frac{p_{i}-\frac{p_{i}}{1-P_G}r(1-P_G)}{p_i} \\
    &= log(1+\frac{r(1-P_G)}{P_G}) \cdot \sum_{i \in G}p_{i} + log(1-r) \cdot  \sum_{i \in R}p_{i} \\
    &= P_G \cdot log(1+\frac{r(1-P_G)}{P_G}) + (1-P_G) \cdot log(1-r) \\
    \end{split}
\end{equation}

Then, we define the multi-objective trade-off analysis function as: 

\begin{equation}
\begin{split}
\mathcal{F}(r) &= \mathcal{T}(r) + \omega \mathcal{W}(r) \\
  &= P_G \cdot log(1+\frac{r(1-P_G)}{P_G}) + (1-P_G) \cdot log(1-r) + 2\omega r (1-P_G)
\end{split}
\end{equation}

where $\omega$ is the weight of detection capability and $\omega > 0$. For generality, we impose no additional restrictions on $\omega$. That is, our following derivation is valid for any $w$. 

The first derivative of $\mathcal{F}$ with respect to $r$ is:

\begin{equation}
\begin{split}
    \frac{\partial \mathcal{F}}{\partial r} &= \frac{(1-P_G)}{1+\frac{r(1-P_G)}{P_G}} - \frac{1-P_G}{1-r} + 2\omega(1-P_G) \\
    &= (1-P_G)(\frac{1}{1+\frac{r(1-P_G)}{P_G}} - \frac{1}{1-r} + 2\omega)
\end{split}
\end{equation}

We only need the sign of the derivative later. To simplify the calculation, we use $S$ to replace the derivative above, as $S$ has the same sign. 

\begin{equation}
S = 2\omega + \frac{1}{1+\frac{r(1-P_G)}{P_G}} - \frac{1}{1-r}
\end{equation}


Next, we need to prove that $\mathcal{F}$ achieves its maximum at $S = 0$. The formula for the first derivative of $S$ with respect to $r$ is: 

\begin{equation}
    \begin{split}
    \frac{\partial S}{\partial r} &= \frac{P_G^2}{(-rP_G + P_G + r)^2} - \frac{P_G}{(-rP_G + P_G + r)^2} - \frac{1}{(r - 1)^2} \\ 
    &= -\frac{P_G(1-P_G)}{(P_G + r - rP_G)^2} - \frac{1}{(1 - r)^2}
    \end{split}
\end{equation}

This derivative is negative, meaning that $S$ is decreasing as $r$ increases. 


\begin{equation}
\lim_{r \to 0} S = 2 \omega > 0
\end{equation}

\begin{equation}
\lim_{r \to 1} S = -\infty < 0
\end{equation}

Since $S$ is positive at $r = 0$ and negative at $r = 1$, by the continuity of $S$ and the intermediate value theorem, there must exist a value $r^*$ between 0 and 1 such that $S(r^*) = 0$. 

By the implicit function theorem, substituting $S = 0$, we obtain the relationship between $r^*$ and $P_G$:

\begin{equation}
\frac{\partial r^{*}}{\partial P_G} = -\frac{\frac{\partial S}{\partial P_G}}{\frac{\partial S}{\partial r^{*}}} = -\frac{\frac{r}{(P_G - r(P_G - 1))^2}}{-\frac{P_G(1-P_G)}{(P_G + r - rP_G)^2} - \frac{1}{(1 - r)^2}} > 0
\end{equation}

This shows that as $P_G$ increases, the optimal $r$ should also increase. We verify the theorem. 

\section{Supplementary Experimental Results}

\subsection{Detailed Experimental Setup}
\label{sec:additional_experimental_setup}

\noindent\textbf{Datasets and Models.} To ensure the reliability, we adapt the configurations provided by MarkLLM~\cite{pan2024markllm}, which currently is the most popular LLM watermarking toolkits. Specifically, for dataset, we utilize 400 samples from the C4 dataset~\cite{raffel2020exploring}. The first 30 tokens of each text serve as prompts to generate new tokens. We set the output length to be at least 200 and at most 230 tokens. We also follow MarkLLM and employ OPT-1.3B, -2.7B and -6.7B~\cite{zhang2022opt} as our models. 

\noindent\textbf{Baselines.} In this paper, we focus exclusively on flexible watermarking methods that do not require training any additional models, as they offer more promising practical applicability. Consequently, we exclude watermarking techniques that necessitate model training, such as SIR~\cite{liu2024sir} and TS~\cite{huo2024token}. The baseline methods include: (1) UnWM, representing the original unwatermarked outputs; (2) KGW~\cite{kirchenbauer2023watermark}, the fundamental method; (3) UW~\cite{hu2024unbiased} and DiPmark~\cite{wu2024resilient}, which implement unbiased watermark techniques; (4) SWEET~\cite{lee2024wrote} and EWD~\cite{lu2024entropy}, both designed for watermarking in low-entropy scenarios. Implementation details can be found in App.~\ref{sec:additional_experimental_setup}. 

\noindent\textbf{Evaluation Metrics.} 
We evaluate MorphMark and baselines in watermark effectiveness and text quality. The evaluation of \textit{effectiveness} focuses on both detectability and robustness. We assess detectability using True positive rate at 1\% false positive rate~(TPR@1\%). We also report the Best F1 Score~(Best F1) to present the highest F1 score achieved with the optimal balance of TPR and FPR during detection. To assess the robustness of watermark methods, we employ the Word-S/30\% attack, which randomly replaces words with synonyms from WordNet~\cite{miller1995wordnet}. We report the TPR@1\% and Best F1 of watermarking methods against the Word-S/30\% attack, denoted as TPR@1\%(Word-S/30\%) and Best F1(Word-S/30\%). From a \textit{text quality} perspective, we evaluate the Perplexity (PPL) of generated texts, computed using LLaMA-2-7B~\cite{touvron2023llama}. 
All experiments are performed on an Ubuntu 18.04 system with an AMD EPYC 7Y83 64-core CPU and a NVIDIA RTX 4090 GPU. 

\noindent\textbf{Implementation Details.} For KGW, SWEET and EWD , and the $\delta$ in their methods is set to 1.25. For SWEET, the entropy threshold is set to 0.9. 
For UW, we use $\gamma$-reweight. For DiPmark, $\alpha$ is set to 0.45. +
For MorphMark$_{linear}$, MorphMark$_{exp}$ and MorphMark$_{log}$, we set $k_{linear}$, $k_{exp}$ and $k_{log}$ to 1.55, 1.30 and 2.15, respectively. $p_0$ in MorphMark is fixed to 0.15. $\epsilon$ is fixed to $10^{-10}.$ 
For all methods, we set the green list ratio to 0.5. 

\subsection{Configuration of Doc-P(GPT-3.5) Attack}
\label{sec:gpt35}

For Doc-P(GPT-3.5) attack, we use the version \text{gpt-3.5-turbo-0125} API. The prompt for paraphrasing is shown in Fig.~\ref{fig:prompt_gpt35}. 

\begin{figure}[htbp]
    \centering
    \begin{tikzpicture}
        \node[draw, rectangle, fill=gray!20, rounded corners] {Please rewrite the following text (Only return the rewritten text): \{Model Output\} };
    \end{tikzpicture}
    \caption{Prompt used in Doc-P(GPT-3.5) paraphrasing attack.}
    \label{fig:prompt_gpt35}
\end{figure}

\subsection{Trade-off Curve Between Watermark Effectiveness and Text Quality}

\begin{figure}[h!]
\centering
    \includegraphics[width=0.40\textwidth]{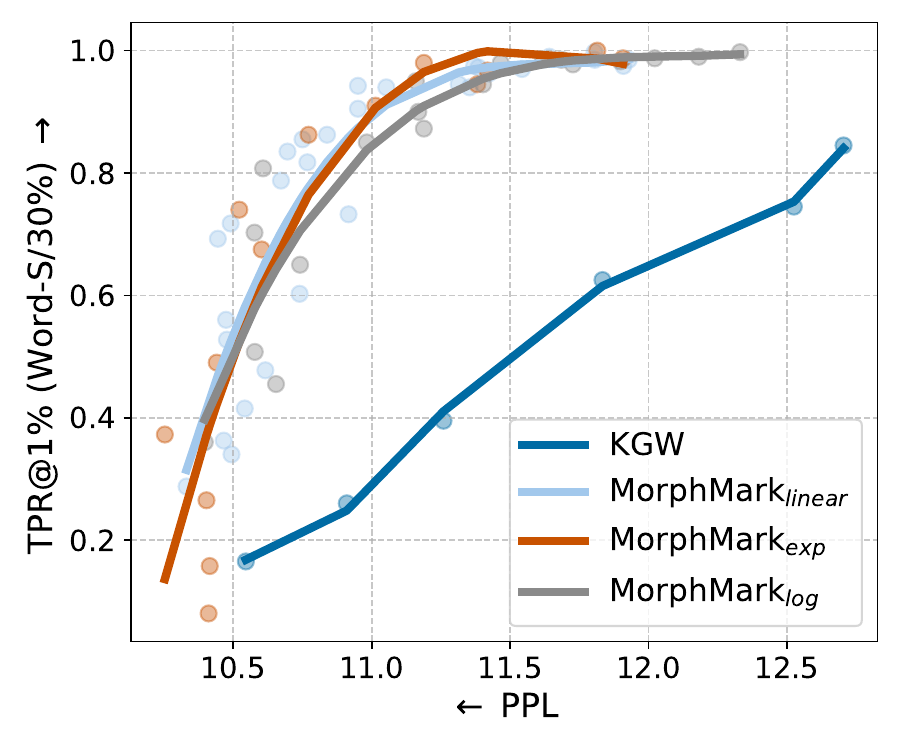}
\caption{Comparing the performance of different watermark methods. We measure watermark effectiveness with TPR@1\%(Word-S/30\%) and text quality with PPL. }
\label{fig:compare_kgw}
\end{figure}

Here, we plot the trade-off curve and compare MorphMark's three varients with KGW. By adjusting $ k_{\text{linear}} $, $ k_{\text{exp}} $, $ k_{\text{log}} $, and $ \delta $, we obtain multiple points, which are visualized in Fig.~\ref{fig:compare_kgw}. From the results, we observe that the MorphMark$_{exp}$ outperforms the MorphMark$_{linear}$, which in turn outperforms the MorphMark$_{log}$. All three methods significantly surpass KGW. 

\subsection{Different Sampling Parameters of More Methods}
\label{sec:sampling_param}

We present more results on different sampling parameters in Tab.~\ref{tab:sampling_param_linear} and Tab.~\ref{tab:sampling_param_log}. 

\begin{table}[h!]
\centering
\scalebox{0.80}{
\begin{tabular}{ccccc}
\Xhline{1.15pt}
\textbf{\begin{tabular}[c]{@{}c@{}}(Temp, TopP)\end{tabular}} & \textbf{UnWM PPL} & \textbf{PPL}     & \textbf{TPR@1\%} & \textbf{\begin{tabular}[c]{@{}c@{}}TPR@1\%$\uparrow$\\ (Word-S/30\%)\end{tabular}} \\ \hline
(0.3, 1.0)         & 4.1308   & 4.6790  & 1.0000  & 0.9025  \\
(0.7, 0.95)        & 5.4809   & 6.1147  & 0.9950  & 0.9325  \\
(0.9, 0.95)        & 7.3829   & 7.9732  & 1.0000  & 0.9325  \\
(1.2, 1.0)         & 15.2175  & 16.3252 & 1.0000  & 0.9625 \\
\Xhline{1.15pt}
\end{tabular}}
\caption{Performance of MorphMark$_{linear}$. }
\label{tab:sampling_param_linear}
\end{table}

\begin{table}[h!]
\centering
\scalebox{0.80}{
\begin{tabular}{ccccc}
\Xhline{1.15pt}
\textbf{\begin{tabular}[c]{@{}c@{}}(Temp, TopP)\end{tabular}} & \textbf{UnWM PPL} & \textbf{PPL}     & \textbf{TPR@1\%} & \textbf{\begin{tabular}[c]{@{}c@{}}TPR@1\%$\uparrow$\\ (Word-S/30\%)\end{tabular}} \\ \hline
(0.3, 1.0)         & 4.1308   & 4.8056  & 0.9925  & 0.9475  \\
(0.7, 0.95)        & 5.4809   & 6.2566  & 0.9950  & 0.9410  \\
(0.9, 0.95)        & 7.3829   & 8.0720  & 1.0000  & 0.9400  \\
(1.2, 1.0)         & 15.2175  & 16.3264 & 1.0000  & 0.9525 \\
\Xhline{1.15pt}
\end{tabular}}
\caption{Performance of MorphMark$_{log}$. }
\label{tab:sampling_param_log}
\end{table}

\subsection{Statistical Distribution of $P_G$ in C4 Dataset}
\label{sec:Distribution_of_P_G}

\begin{figure*}[t!]
\centering
    \subfigure[TruthfulQA]{
        \includegraphics[width=0.2250\textwidth]{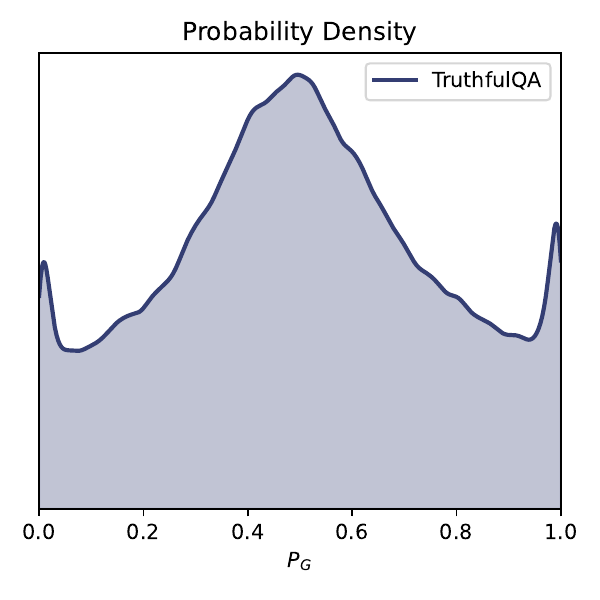}
        \label{fig:TruthfulQA}
    }\hspace{-2.5mm}
    \subfigure[SQuAD]{
        \includegraphics[width=0.2250\textwidth]{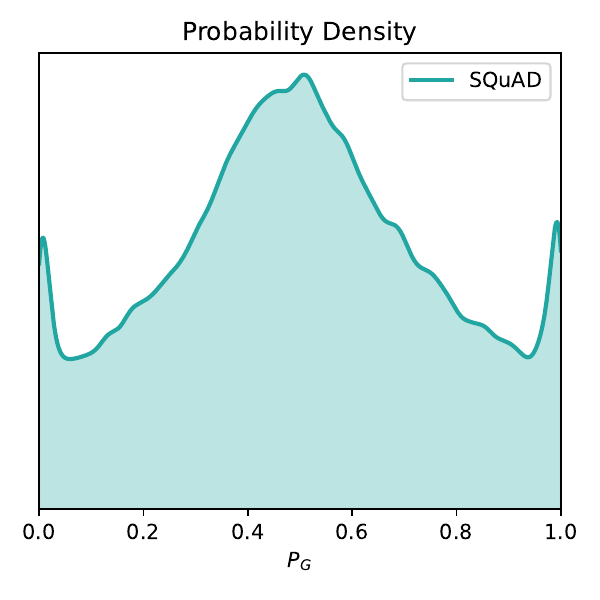}
        \label{fig:SQuAD}
    }\hspace{-2.5mm}
    \subfigure[GSM8K]{
        \includegraphics[width=0.2250\textwidth]{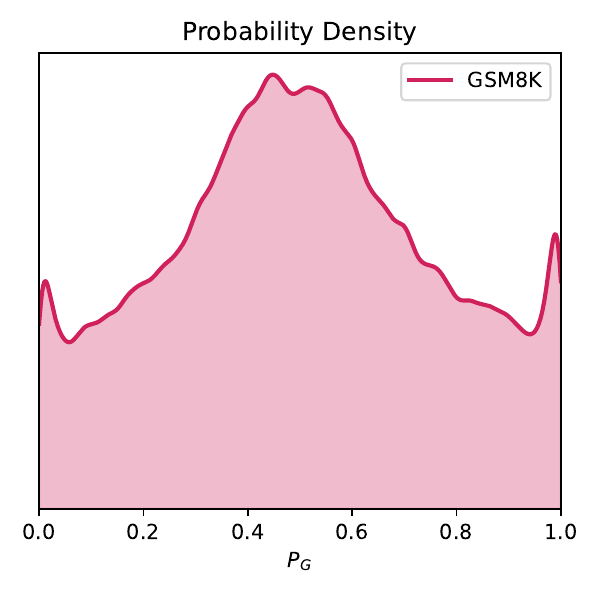}
        \label{fig:GSM8K}
    }\hspace{-2.5mm}
    \subfigure[MBPP]{
        \includegraphics[width=0.2250\textwidth]{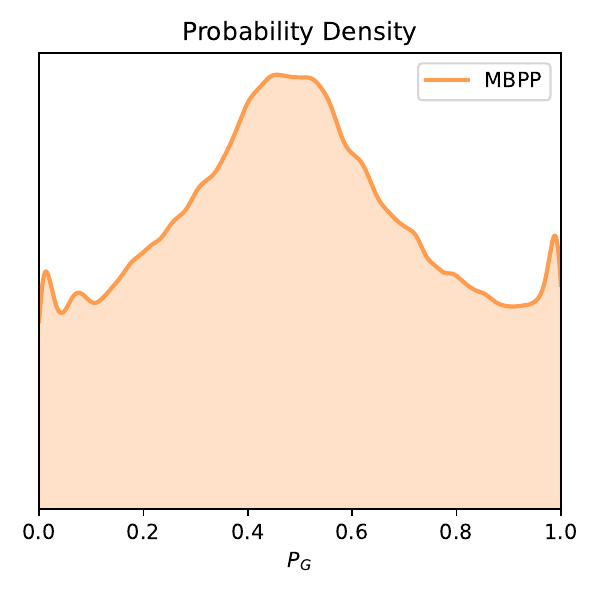}
        \label{fig:MBPP}
    }
\caption{Statistical Distribution of $P_G$. }
\label{fig:PG_dis}
\end{figure*}

Before, we discuss an extreme case of code generation which make MorphMark low effectiveness. To further explore the occurrence of extreme cases, we use the questions in four popular benchmarks, i.e., TruthfulQA~\cite{lin2021truthfulqa}, SQuAD~\cite{rajpurkar2016squad, rajpurkar2018know}, GSM8K~\cite{cobbe2021gsm8k} and MBPP~\cite{austin2021program}. For each dataset, we randomly sample 400 questions and subsequently analyze the resulting $P_G$ distribution, as shown in Fig.~\ref{fig:PG_dis}. These empirical results indicate that the $P_G$ distribution is generally broad. Although the code generation dataset MBPP exhibits more values close to 0 and 1 compared to other datasets, its overall distribution remains broad. This observation suggests that MorphMark is effective in a wide range of scenarios. 


\subsection{Comparative Analysis of Z-Scores with Low-Entropy Watermarking Methods}
\label{sec:analysis_low_entropy}

To further understand the performance of different watermarking methods under low-entropy scenarios, we conduct a comparative analysis of z-scores for MorphMark and EWD on the HumanEval dataset, using the same experimental settings as in the previous section. The results are summarized in Table~\ref{tab:zscore_comparison}.

\begin{table}[ht]
    \centering
    \begin{tabular}{lccc}
        \toprule
        Method & Mean Z-score (Watermarked)$\uparrow$ & Mean Z-score (UnWM) & TPR@1$\uparrow$ \\
        \midrule
        SWEET        & 2.9027 & \textbf{-0.0252} & 0.7073 \\
        EWD            & 3.3399 & -0.0499 & 0.8537 \\
        MorphMark          & 3.8824 & 0.1721  & 0.8780 \\
        MorphMark + EWD           & \textbf{3.9977} & -0.0499 & \textbf{0.9207} \\
        \bottomrule
    \end{tabular}
    \caption{Comparative analysis of z-scores for watermarked and non-watermarked text using different watermarking methods.}
    \label{tab:zscore_comparison}
\end{table}

From these results, several key insights can be observed:

\begin{itemize}
    \item \textbf{SWEET} reduces the z-score of non-watermarked text. However, it does not achieve a higher z-score for watermarked text compared to EWD and MorphMark, nor does it demonstrate superior detectability (TPR@1) over these methods.

    \item \textbf{EWD} increases the z-score of watermarked text while decreasing that of non-watermarked text. This is due to its adaptive strategy, which selects a subset of high-entropy text for watermarking, resulting in a more pronounced watermark signal. Moreover, EWD lowers the z-score of non-watermarked text by reducing the influence of low-entropy tokens, which are generally less suitable for watermarking. This aligns with the original statement in the EWD paper~\cite{ewd}: ``Our EWD method results in overall higher z-scores for watermarked texts and slightly lower z-scores for human texts.''

    \item \textbf{MorphMark} increases the z-score of watermarked text while maintaining a relatively stable z-score for non-watermarked text. This indicates that MorphMark effectively strengthens the watermark signal without significantly degrading the text quality.

    \item \textbf{MorphMark + EWD} combines the strengths of both methods. The combination achieves the largest z-score gap between watermarked and non-watermarked texts, leading to the highest detectability (TPR@1). This result demonstrates that integrating MorphMark's adaptive watermarking with EWD's entropy-based detection further enhances watermark robustness.
\end{itemize}

These observations provide a clearer understanding of the advantages and limitations of each method, highlighting the effectiveness of MorphMark, especially when combined with EWD. 

\section{Full Related Work}
\label{sec:full_related_work}

\noindent \textbf{Watermarking in the Era of LLMs} Modern watermarking techniques for large language models (LLMs) differ significantly from earlier backdoor-based approaches, primarily due to the high costs of training such models. Instead of embedding watermarks during training, contemporary methods apply them during the sampling phase of text generation. The pioneering method in this space is KGW~\cite{kirchenbauer2023watermark}, which utilizes a user-defined key and the previous token as a random seed to split the vocabulary into "green" and "red" lists. The model then increases the probabilities of green-list tokens to embed the watermark. Since KGW's introduction, numerous techniques have sought to enhance its performance from various perspectives. 

\noindent \textbf{Unbiased Watermarking} Unbiased watermarking ensures that the expected token distribution under watermarking remains identical to the original. The first method to achieve this, EXP, is highly computationally expensive. For example, \citet{wu2024resilient} reports that EXP can require up to 500 times the generation time of KGW. More efficient alternatives, such as UW and DipMark, leverage inverse sampling and permutation-based reweighting to strike a balance between detection efficacy and text quality. However, their robustness has yet to be thoroughly validated. 

\noindent \textbf{Semantics-Based Watermarking} A growing body of research~\cite{ren2024robust, liu2024sir, he2024can, guo2024context} has explored the use of semantic information, rather than previous tokens, as keys for embedding watermarks. This approach enhances robustness without increasing watermark strength, thereby preserving text quality. However, many of these methods require auxiliary models, reducing their flexibility. Among them, SIR~\cite{liu2024sir} demonstrated the strongest performance in the MarkLLM benchmark, making it a key baseline in our study. 

\noindent \textbf{Low-Entropy Watermarking} Low-entropy contexts involve highly deterministic token generation—e.g., completing The quick brown fox jumps over a lazy, where dog is the most probable next token. In such cases, watermarking can degrade text quality. Methods like SWEET~\cite{lee2024wrote} and ATW~\cite{liu2024adaptive} mitigate this by setting entropy thresholds, embedding watermarks only when token uncertainty is sufficiently high. EWD~\cite{lu2024entropy} takes a different approach, maintaining the KGW framework but assigning higher detection weights to high-entropy tokens. However, these techniques often require access to the original model during detection, limiting practicality—especially ATW, which relies on three auxiliary models, making both watermarking and detection computationally expensive. 

\noindent \textbf{Other Watermarking Techniques} Unigram~\cite{zhaoprovable} improves robustness by using a fixed vocabulary partition instead of dynamically adjusting token probabilities based on prior tokens. However, this fixed division is vulnerable to watermark extraction techniques~\cite{jovanovicwatermark}, making it impractical for real-world applications. 
TS~\cite{huo2024token} converts the hyperparameters in KGW into two neural networks and designs a loss function for training to enhance both watermark effectiveness and text quality. However, this approach not only lacks interpretability, but also requires retraining a new watermark parameter neural network for every new model. More importantly, in practical applications, the watermark strength is difficult to control manually and becomes unpredictable due to its training-based nature.

\section{Further Analysis of the Connection Between Theory and Practical Implementation}

This section clarifies the relationship between our theoretical derivations and practical implementation. Our theoretical analysis establishes the existence of an optimal solution $r^*$ that maximizes the multi-objective function. However, this solution lacks a closed-form expression due to the complexity of the function $S(r)$, which depends on two parameters: $P_G$ and $\omega$. Specifically, the solution is implicitly defined as:
\begin{equation}
    S(r^*) = 0 \implies r^* = \varepsilon(\omega, P_G).
\end{equation}
The function $\varepsilon$ is implicitly defined and cannot be directly expressed, making a closed-form solution intractable.

\noindent Although a numerical approximation of $r^*$ is theoretically feasible, we choose not to pursue this approach. The primary reason is the introduction of human bias through the parameter $\omega$, which reflects user preferences for watermark effectiveness. Setting this parameter manually would undermine the objectivity of the method. Moreover, calculating a numerical solution for each combination of $P_G$ and $\omega$ is computationally expensive, especially given the dynamic nature of watermarking scenarios.

\noindent To maintain both theoretical rigor and practical efficiency, we bypass the complexity associated with $\omega$ by leveraging the positive relationship between $r^*$ and $P_G$. Our empirical results demonstrate that this approximation achieves a strong balance between watermark effectiveness and text quality, consistent with the theoretical insights.

\noindent In summary, while our practical implementation does not directly solve for $r^*$, it effectively realizes the theoretical principles, avoiding unnecessary complexity and ensuring robust performance.

\end{document}